\documentclass{conm-p-l}
\copyrightinfo{2003}{American Mathematical Society}
\usepackage{amsmath}

\newtheorem{theorem}{Theorem}[section]

\theoremstyle{definition}

\theoremstyle{remark}
\newtheorem{remark}[theorem]{Remark}

\numberwithin{equation}{section}

\newcommand{\abs}[1]{\lvert#1\rvert}

\makeatletter
\def \@cflb {%
    \let\@elt\@comflelt
    \setbox\@tempboxa \vbox{}%
    \@botlist
    \setbox\@outputbox \vbox{%
                             \unvbox\@outputbox
                             \vskip \textfloatsep
                             \botfigrule
                             \unvbox\@tempboxa
                             \vskip -\floatsep
                             }%
    \let\@elt\relax
    \xdef\@freelist{\@freelist\@botlist}%
    \global \let \@botlist\@empty
}
\makeatother

\begin{document}

\def\Cx{{\bf C}} \def\Z{{\bf Z}} \def\x{{\bf x}} \def\o{{\bf o}}
\def\del{\partial} \def\Lap{\bigtriangleup} \def\Tr{{\rm Tr}}
\def\^{\wedge} \def\goinf{\rightarrow\infty} \def\goes{\rightarrow}
\def\bm{\boldmath} \def\-{{-1}} \def\inv{^{-1}} \def\sqr{^{1/2}}
\def\isqr{^{-1/2}}

\def\reff#1{(\ref{#1})}
\def\vb#1{{\partial \over \partial #1}} 
\def\vbrow#1{{\partial/\partial #1}} 
\def\Del#1#2{{\partial #1 \over \partial #2}}
\def\Dell#1#2{{\partial^2 #1 \over \partial {#2}^2}} \def\Dif#1#2{{d
    #1 \over d #2}} \def\Lie#1{ \mathcal{L}_{#1} } \def\diag#1{{\rm
    diag}(#1)} \def\abs#1{\left | #1 \right |} \def\rcp#1{{1\over #1}}
\def\paren#1{\left( #1 \right)} \def\brace#1{\left\{ #1 \right\}}
\def\bra#1{\left[ #1 \right]} \def\angl#1{\left\langle #1
  \right\rangle} \def\vector#1#2#3{\paren{\begin{array}{c} #1 \\ #2 \\
      #3 \end{array}}} \def\svector#1#2{\paren{\begin{array}{c} #1 \\
      #2 \end{array}}}

\def\GL#1{{\rm GL}(#1)} \def\SL#1{{\rm SL}(#1)} \def\PSL#1{{\rm
    PSL}(#1)} \def\O#1{{\rm O}(#1)} \def\SO#1{{\rm SO}(#1)}
\def\IO#1{{\rm IO}(#1)} \def\ISO#1{{\rm ISO}(#1)} \def\U#1{{\rm
    U}(#1)} \def\SU#1{{\rm SU}(#1)}

\def\diffeos{diffeomorphisms} \def\diffeo{diffeomorphism}
\def\Teich{{Teichm\"{u}ller}} \def\Poin{{Poincar\'{e}}}

\def\Gam{\mbox{$\Gamma$}} \def\d{{d}} \def\VII#1{\mbox{VII${}_{#1}$}}
\def\VI#1{\mbox{VI${}_{#1}$}} \def\Nil{{\rm Nil}} \def\Sol{{\rm Sol}}
\def\Isom{{\mathrm{Isom}}}

\def\hh{{h}}
\def\ggg{{\rm g}} \def\uh#1#2{\hh^{#1#2}} \def\dh#1#2{\hh_{#1#2}}
\def\mh#1#2{\hh^{#1}{}_{#2}} \def\ug#1#2{\ggg^{#1#2}}
\def\dg#1#2{\ggg_{#1#2}} \def\uug#1#2{\tilde{\ggg}^{#1#2}}
\def\udg#1#2{\tilde{\ggg}_{#1#2}} \def\udh#1#2{\tilde{\hh}_{#1#2}}

\def\c#1{\chi_{#1}} \def\cc#1#2{\chi_{#1}{}^{#2}} \def\uc#1{\chi^{#1}}
\def\s#1{\sigma^{#1}} \def\ss#1#2{\sigma^{#1}{}_{#2}}
\def\om#1#2#3{\omega_{#1#2}{}^{#3}} \def\omd#1#2#3{\omega_{#1#2#3}}
\def\CC{C} \def\C#1#2#3{\CC^{#1}{}_{#2#3}} \def\Sig#1{\Sigma^{#1}}
\def\Sigg#1#2{\Sigma^{#1}{}_{#2}} \def\Chi#1{X_{#1}}
\def\Chii#1#2{X_{#1}{}^{#2}} \def\dug#1#2{g_{#1}{}^{#2}}
\def\dgm#1#2{\gamma_{#1#2}} \def\ugm#1#2{\gamma^{#1#2}}
\def\covsset{\mathcal{S}_2} \def\consset{\mathcal{S}^2} \def\X{X}
\def\stdh{h_{\mathrm{s}}{}} \def\dd#1#2{\frac{\del^2{#1}}{\del {#2}^2}}

\def\wa{\!\!\!\!&=&\!\!\!\!}  \def\wb{\!\!\!\!&\equiv &\!\!\!\!}
\def\ws{&=} \def\nd{\noindent} \def\D{{\mathcal{D}}}
\def\tr{\mathrm{tr}} \def\nnpar{\nonumber \\} \def\xnnpar{\\}

\def\proofmark{\textit{Proof}: } \def\defmark{{\bf
    Definition}\hspace{1em}} \def\defsmark{{\bf
    Definitions}\hspace{1em}} \def\notemark{{\bf Note}\hspace{1em}}
\def\remarkmark{{\bf Remark}\hspace{1em}} \def\conventionmark{{\bf
    Convention}\hspace{1em}}
\def\endofproofmark{\hfill\rule{.5em}{.5em}}

\def\hD{{\hat D}} \def\E#1#2#3{(E_{#1})_{#2#3}}
\def\uE#1#2#3{(E_{#1})^{#2#3}} \def\dpi#1{\pi_{#1}} \def\dq#1{q_{#1}}
\def\lmx{\sqrt{\frac{\lambda^2+2}{2}}}

\def\TT{\textsc{tt}}  \def\Q#1{Q_\mathrm{#1}} \def\tm{t_-}
\def\tp{t_+} \def\nt{{(0)}} \def\Mes#1{M_\mathrm{#1}}
\def\R{\mathbf{R}} \def\teiB{\delta} \def\barphi{{\bar\phi}} \def\X{X}

\title{Linear perturbations of spatially locally homogeneous
  spacetimes}

\author{Masayuki Tanimoto}
\address{Department of Physics, Yale University, New Haven, CT 06511}
\curraddr{Max-Planck-Institut f\"ur Gravitationsphysik, Golm 14476,
  Germany}
\thanks{The author was supported by the Japan Society for the
  Promotion of Science (JSPS). The author is grateful to Vincent
  Moncrief and Katsuhito Yasuno for collaborations related to this
  work and helpful discussions.}

 \date{March,
  2003.}


\begin{abstract}
  Methods and properties regarding the linear perturbations are
  discussed for some spatially closed (vacuum) solutions of Einstein's
  equation. The main focus is on two kinds of spatially locally
  homogeneous solution; one is the Bianchi III (Thurston's
  $H^2\times\mathbf{R}$) type, while the other is the Bianchi II
  (Thurston's Nil) type. With a brief summary of previous results on
  the Bianchi III perturbations, asymptotic solutions for the
  gauge-invariant variables for the Bianchi III are shown, with which
  (in)stability of the background solution is also examined. The issue
  of linear stability for a Bianchi II solution is still an open
  problem. To approach it, appropriate eigenfunctions are presented
  for an explicitly compactified Bianchi II manifold and based on
  that, some field equations on the Bianchi II background spacetime
  are studied. Differences between perturbation analyses for Bianchi
  class B (to which Bianchi III belongs) and class A (to which Bianchi
  II belongs) are stressed for an intention to be helpful for
  applications to other models.
\end{abstract}





\maketitle
\section{Introduction}

A classification of three dimensional homogeneous (Riemannian)
manifolds is known in relativity as the Bianchi classification, which
is a classification of the three dimensional simply transitive Lie
groups by isomorphism. A four dimensional spacetime $(X\times\R,\dg
ab)$ is said to be spatially homogeneous if a Bianchi group $G$ acts
spacelike and makes the metric $\dg ab$ invariant. Cosmological models
based on such a spacetime structure, especially with isotropy, have
long been a basis for cosmology, and (in case of being anisotropic)
have recently been subjects for more mathematical study in relativity
(e.g.\cite{WE}). Departure from a homogeneity however belongs to a
very recent development in the field (except for linear perturbations
of rather special homogeneous solutions such as those with isotropy or
ones that are not isotropic but can become so in a limit). In
particular, whether the solution is stable or not is unknown for
almost all such solutions.

This article discusses linear perturbations of two Bianchi type
solutions; one is Bianchi III, the other is Bianchi II. The spatial
manifold of each solution is assumed to be \textit{closed} (meaning
`compact without boundary'), introducing an appropriate
compactification. (As a result, the solution becomes \textit{locally
homogeneous}.) One of the common features of these two models is that
both of the spatial manifolds belong to so-called Yamabe type $-1$,
which is of special interest in relativity \cite{FM}. One of the
biggest differences is that while Bianchi III belongs to so-called
class B, Bianchi II belongs to class A. (The class A or B is defined
with respect to whether the trace of the structure constant of a
Bianchi group vanishes or not. \cite{EM}) In fact, we will see some of
the largest differences between Bianchi III and II amount to those
between class A and B. Therefore, in particular, the basic methods for
Bianchi II will be applicable to other class A Bianchi models,
too. (Difficulties in Bianchi II will also be common for all Bianchi A
models.)

We deal with the Bianchi III perturbations first. In this case, we
have almost complete results. The technical details of fundamental
analysis of Bianchi III vacuum perturbations, including the derivation
of the wave equations for the gauge-invariant variables, have already
been given in \cite{TMY} and is also briefly summarized in subsections
\ref{ssec:III-1} to \ref{ssec:III-3} of the present article. The
asymptotic stability of the perturbations has been discussed in
\cite{YT} in a framework of dynamical system approach, where one does
not have to appeal to an exact background solution. In the present
article we discuss the asymptotic stability from a different point of
view in subsections \ref{ssec:III-4} and \ref{ssec:III-5}, using the
exact background solution. The issue of stability of the Bianchi II
solution is on the other hand still an open question. In this article
we discuss properties of some simpler field equations on that
background in the subsequent section, stressing differences from the
Bianchi III background.

\section{Bianchi III vacuum perturbations}

\subsection{The background solution}
\label{ssec:III-1}

First, let us describe the background geometry briefly.

The Bianchi III algebra is generated by three vectors $\xi_I$
($I=1\sim3$) such that the commutation relation is given by
\begin{equation}
  [\xi_1,\xi_2]=\xi_1,\quad [\xi_2,\xi_3]=[\xi_3,\xi_1]=0.
\end{equation}
Using coordinates, they can be represented as
\begin{equation}
  \xi_1=\vb x,\quad \xi_2=x\vb x+y\vb y,\quad \xi_3=\vb z,
\end{equation}
We will be able to think of these vectors as Killing vectors on our
background spacetime. One can also consider invariant vectors $\chi_I$
and 1-forms $\s I$ defined by
\begin{equation}
  \Lie{\xi_I}\chi_J=0,\; \mathrm{and}\quad \Lie{\xi_I}\s J=0,
\end{equation}
where $\Lie{\xi_I}$ denotes the Lie derivative with respect to the
vector ${\xi_I}$. Suppose $\chi_I$ and $\s I$ are dual to each other;
$\angl{\chi_I,\s J}=\delta_I^J$. Then, using the coordinates they are
represented as
\begin{equation}
  \begin{split}
    \chi_1=y\vb x,\quad \chi_2=y\vb y,\quad \chi_3=\vb z, \\
    \s1=\frac{dx}y,\quad \s2=\frac{dy}y,\quad \s3={dz}.
  \end{split}
\end{equation}

For convenience we define a standard 2-dimensional hyperbolic
metric
\begin{equation}
  \tilde h\equiv\s1\otimes\s1+\s2\otimes\s2,
\end{equation}
and a standard 1-dimensional Euclid metric
\begin{equation}
  \tilde l\equiv \s3\otimes\s3.
\end{equation}

Then, the (universal covering) vacuum solution we consider is given by
\begin{equation}
  \tilde \ggg=-N(t)^2dt^2+q_1(t)\tilde h+q_2(t)\tilde l,
\end{equation}
where the two scale functions $q_1(t)$ and $q_2(t)$ and the lapse
function $N(t)^2$ are explicitly given by
\begin{equation}
  q_1(t)=(t+k)^2,\quad q_2(t)=(t-k)/(t+k),\quad N(t)^2=1/q_2(t),
\end{equation}
where $k$ is a real parameter.

This type of solution is called \textit{locally rotationally symmetric
  (LRS)}, because, due to the hyperbolic metric $\tilde h$ the
spatial metric $\tilde q\equiv q_1\tilde h+q_2\tilde l$ has a fourth
Killing vector (with degenerate points). We consider only this type of
metric, because otherwise one cannot compactify the spatial manifold.

We can express the compactified solution as follows:
\begin{equation}
  (M\times\R,\ggg)=(\R^4,\tilde \ggg)/\Gamma,
\end{equation}
where the spatial manifold $M$ is supposed to be the direct product
$\Sigma_g\times S^1$, where $\Sigma_g$ ($g>1$) is a higher genus
surface with genus $g$ while is $S^1$ the circle. For convenience, we
call $\Sigma_g$ \textit{the base} and $S^1$ \textit{the
  fiber}. $\Gamma\subset\Isom(\R^4,\tilde \ggg)$ is a discrete
subgroup of the isometry group of $(\R^4,\tilde \ggg)$, which acts on
$(\R^4,\tilde \ggg)$ from the left. We assume that $\Gamma$ is of
\textit{orthogonal} type\cite{TMY}, which means that (the action of)
$\Gamma$ preserves the foliation by $H^2$ of the universal covering
manifold.  In this case one can write the spacetime metric $\ggg$ for
the spatially compactified manifold in almost the same form as $\tilde
\ggg$:
\begin{equation}
  \ggg=-N(t)^2dt^2+q_1(t) h+q_2(t) l,
\end{equation}
where $h$ is the \textit{hyperbolic} metric on $\Sigma_g$ induced from
the universal cover metric $\tilde q$, while $l$ is the Euclid metric
on $S^1$ induced from the same metric.

\subsection{Eigenfunctions, vectors, and symmetric tensors}
\label{ssec:III-2}

We need to find complete sets of eigenfunctions, (eigen)vectors, and
(eigen)symmetric tensors for the Laplacian so that we can mode-expand
the field variables on our background. Thanks to the fact that our
spatial manifold has the orthogonal direct product property, we can
reduce this task to those for the base and fiber. Namely, once one
finds eigenfunctions, vectors, and symmetric tensors on both the fiber
and the base, the remaining task is just to make appropriate products.
For details, see \cite{TMY}. Let us in the following recall some basic
definitions.

The most fundamental are the eigenfunctions on the fiber and the
base. Let $c_m$ be a real eigenfunction on the fiber, corresponding to
eigenvalue $m$;
\begin{equation}
  (\c3)^2c_m=-m^2 c_m.
\end{equation}
We call $m$ \textit{the fiber eigenvalue}. ($\c3=\del_z$ is the
derivative operator along the fiber.) We also define another real
eigenfunction $\bar c_m$ by
\begin{equation}
  \c3 c_m=-m \bar c_m, \quad \c3 \bar c_m=m c_m.
\end{equation}
Similarly, let $\hat S_\lambda$ be a real eigenfunction on the base,
corresponding to eigenvalue $\lambda$;
\begin{equation}
  \Lap_h\hat S_\lambda=-\lambda^2 \hat S_\lambda.
\end{equation}
We call $\lambda$ \textit{the base eigenvalue}. ($\Lap_h$ is the
Laplacian with respect to the metric $h$.) The mode function on
the spatial manifold $M=\Sigma_g\times S^1$ is given by the product
$ S_{m,\lambda}\equiv c_m \hat S_\lambda$;
\begin{equation}
  \Lap_q S_{m,\lambda}=-(q_1\inv\lambda^2+q_2\inv m^2)S_{m,\lambda},
\end{equation}
where $q=q_1h+q_2l$ is the spatial part of the metric $\ggg$.

The mode vectors and symmetric tensors on $(M,q)$ are categorized,
according to those for the base $(\Sigma_g,h)$, into four kinds; the
\textit{even} type, the \textit{odd} type, the \textit{harmonic} type,
and the \textit{transverse-traceless} (\TT) type. The meaning of this
is as follows.

The even vectors (or 1-forms) $\hat S_a$ on the base are equivalent to
the exact (gradient) 1-forms on the base, $\hat S_a=\hat D_a \hat S$,
where $\hat D_a$ is the covariant derivative operator on the
base. $\hat S$ is generally a function on the base, but it is supposed
to be the mode function $\hat S=\hat S_\lambda$ when thinking of $\hat
S_a$ as a mode vector on the base.  Similarly, the odd 1-forms $\hat
V_a$ are equivalent to the (Hodge-)dual exact 1-forms, $\hat
V_a=\varepsilon_a{}^b\hat D_b \hat S$, where
$\varepsilon_{ab}=2\ss1{[a}\ss2{b]}$ is the area 2-form on the
base. (To raise indices, use $\uh ab$, the inverse of $\dh ab$.) The
harmonic vectors $\hat U_a$ are the Hodge harmonic vectors defined by
$d\hat U=\delta\hat U=0$, or equivalently $\hat D_{[a}\hat U_{b]}=\hat
D^a\hat U_a=0$.  The number of the independent harmonic one-forms is
equivalent to the 1st-Betti number of the surface, which is given by
$2g$ in case of the higher genus surface.  The Hodge-decomposition
tells us that an arbitrary 1-form on the base can be uniquely
decomposed using these even, odd, and harmonic (mode) 1-forms.

The even symmetric tensors are symmetric tensors made from the even
vectors or the metric $\dh ab$ and the mode functions. (The mode
functions themselves are supposed to be of even type.)  We define the
\textit{even trace part} by $\hat Sh_{ab}$, and the \textit{even
  traceless part} $\hat S_{ab}$ as the traceless part of the gradient
of $\hat S_a$. Similarly, we define the \textit{odd symmetric tensors}
$\hat V_{ab}$ and the \textit{harmonic symmetric tensors} $\hat
U_{ab}$ by the (symmetric) gradient of $\hat V_a$ and $\hat U_a$,
respectively. (They are automatically traceless.) From York's
decomposition \cite{Yo}, we know that it is necessary for completeness
to consider the \textit{transverse-traceless (\TT) tensors} $\hat
W_{ab}$, which are defined by $\hat D^b\hat W_{ab}=h^{ab}\hat
W_{ab}=0$.  The number of independent \TT\ tensors on the higher genus
hyperbolic surface is the same as the dimension of the so-called
\Teich\ space, which is equal to $6g-6$.

The symmetric mode tensors on the spatial manifold $M$ have, as a
result, nine kinds, which are denoted as $(E_i)_{ab}$ ($i=1,\cdots,
9)$. Four of them are of the even kind, which are defined as
$\E1ab=c_m\hat Sh_{ab}$, $\E2ab=c_m\hat S_{ab}$, $\E3ab=c_m\hat S
l_{ab}$, and $\E4ab=2\bar c_m\hat S_{(a}\ss3{b)}$.  (We omit the
subscript $\lambda$ like $\hat S=\hat S_\lambda$.)  Similarly, the odd
symmetric mode tensors are defined as $\E5ab=c_m\hat V_{ab}$, and
$\E6ab=2\bar c_m\hat V_{(a}\ss3{b)}$.  The harmonic symmetric mode
tensors are defined as $\E7ab=c_m\hat U_{ab}$, and $\E8ab=2\bar
c_m\hat U_{(a}\ss3{b)}$. Finally,  the \TT\ symmetric mode tensors are
defined as $\E9ab=c_m\hat W_{ab}$. These are orthogonal to each other
with respect to the standard $L^2$-norm. (The mode vectors on $M$ are
also defined in a similar way.)

\subsection{Vacuum perturbations and wave equations}
\label{ssec:III-3}

Using the symmetric mode tensors defined above, we can expand the
first variation $\delta q_{ab}$ of the spatial metric $q_{ab}$ as
follows. 
\begin{equation}
  \delta q_{ab}=\sum \gamma^i\E iab,
\end{equation}
where the sum is taken over $i=1,\cdots,9$, as well as all possible
eigenvalues $\lambda$ and $m$. The coefficients $\gamma^i=\gamma^i(t)$
are functions of time that are supposed to be perturbation variables.

An important fundamental result that can be confirmed by direct
computations is about how the perturbations decouple, which is stated
as follows. \textit{For given base eigenvalue $\lambda$ and the fiber
  eigenvalue $m$, the set of the even variables, the odd set, the
  harmonic set and \TT\ set evolve independently from each other.} Note
that this statement says two kinds of decoupling; first of all,
different eigenmodes decouple from each other, and the four kinds of
set of variables decouple from each other. Due to this decoupling
property, we can deal with the perturbations one by one for each one
of the four kinds for each mode.

The perturbation variables change their form by an infinitesimal
\diffeo\ acting on the manifold. Certain combinations of the variables
however remain invariant. Such a combination is known as the
\textit{gauge invariant variable}. In general, one of the central
issues in perturbation analysis is to find these gauge invariant
variables and determine the dynamics of them. Our choice of the
gauge-invariant variables is as follows:
\begin{equation}
  \label{eq:defQs}
  \begin{split}
    Q_{\mathrm{E}} &=
    -\Delta_1\gamma^1-\Delta_2\gamma^2+\gamma^3-(2m/\lambda)\gamma^4, \\
    Q_{\mathrm{O}} &= (\nu/\sqrt2)\gamma^5+\gamma^6, \\
    Q_{\mathrm{H}} &= (m/2)\gamma^7+\gamma^8, \\
    Q_{\mathrm{T}} &= \gamma^9,
  \end{split}
\end{equation}
where $\Delta_1\equiv \dot q_2/\dot q_1$, $\Delta_2\equiv
(\lambda/\sqrt{2(\lambda^2+2)})(\Delta_1+2m^2/\lambda^2)$, and
$\nu\equiv m/\sqrt{\lambda^2+2}$. The subscripts E, O, H, and T are
attached to express it is for the even, odd, harmonic or \TT\
perturbation, respectively.

Getting wave equations (equivalent to the linearized Einstein
equation) is a substantial part of the work in the study of
perturbations. While we show only the results of it in the following,
it is worth mentioning that the actual computations were done in the
Hamiltonian formalism, using so-called the method of generating
function to find a desired canonical transformation that is necessary
to split the system into gauge-dependent and independent parts. After
lengthy computations \cite{TMY}, we get the wave equations for the
gauge invariant variables, which are second order ODEs of the form
\begin{equation}
  \label{eq:waveeqs}
  {\ddot Q}_\mathrm{A}+
  f_\mathrm{A}(t){\dot Q}_\mathrm{A}+
  g_\mathrm{A}(t)\Q{A}=0. \quad
  (\mathrm{A}=\mathrm{E},\mathrm{O},\mathrm{H}, \text{ or } \mathrm{T})
\end{equation}
Each coefficient function $f_\mathrm{A}(t)$ or $g_\mathrm{A}(t)$ are
given as follows:

\textbf{(i)} For the even perturbations:
\begin{equation}
  f_\mathrm{E} =-2\bigg({\frac{(t-2k)}{t_+t_-}}-\frac{X}{Z}\bigg), \quad
  g_\mathrm{E} = m^2\,\frac{t_+^2}{t_-^2}+\lambda^2\rcp{t_+t_-}
  +\frac{Y}{Z},
\end{equation}
where
\begin{equation*}
\begin{split}
X & \equiv 
   8\,m^2(t-2k){\tp}^2+
   2\,{{\lambda}}^2\,m^2( 2 t-k ) \,{\tp}^2 +
   2{{\lambda}}^2(2\,t - 3\,k )  +
   {{{\lambda}}^4(2\,t-k )t}/{\tp},
   \\
Y & \equiv 
    16\,m^4\,{{t_+}}^4
    -4\,k\,m^2\,{{t_+}} \left[
      {{\lambda}}^2\,{(t-4k)}/{\tm} - 8 \right]
    +2\,k^2\,{{\lambda}}^4\,{(2\,t +k)}/{(\tm\tp^{2})},
    \\
Z & \equiv 
       4\,m^4\,{{t_+}}^6
      + 8\,m^2\,{{t_-}}\,{{t_+}}^3
      + 4\,{{\lambda}}^2\,m^2\,t\,{{t_+}}^3
      + 2\,{{\lambda}}^2\,{{t_-}}\,{{t_+}}
      + {{\lambda}}^4\,t^2.
\end{split}
\end{equation*}

\textbf{(ii)} For the odd and harmonic perturbations ($\lambda\neq0$
for the odd, $\lambda=0$ for the harmonic)
\begin{equation}
 f_\mathrm{O} =-\frac{2\nu^2(t-2k)\tp}{\tm u}, \quad
 g_\mathrm{O} =
    (\lambda^2 +2)\frac{u}{\tm^2}
    +\frac{2(t-3k)}{\tp^2 \tm}
    -\frac{4(t-2k)}{\tp^3u},
\end{equation}
where $u\equiv \nu^2t_+^2+{t_-}{t_+\inv}$.

\textbf{(iii)} For the \TT\ perturbations
\begin{equation}
  f_\mathrm{T} = -\frac{2(t-2k)}{\tp\tm},\quad
  g_\mathrm{T} = m^2\frac{\tp^2}{\tm^2}+\frac{2(t-3k)}{\tp^2\tm}.
\end{equation}

\subsection{Asymptotic solutions}
\label{ssec:III-4}

Let us find future asymptotic solutions to discuss the future
stability. A lot is known about how we can find asymptotic solutions
for linear ODEs. The following simple result, taken from standard
texts, is sufficient for our purpose.
\begin{theorem}[e.g., \cite{CH}]
  \label{th:1}
  Consider the following second order ODE
  \begin{equation}
    \label{eq:ode1}
    X''+f(s)X=0,
  \end{equation}
where $f(s)$ is a function of $s$ that approaches a constant
\begin{equation}
  \lim_{s\goes\infty}f(s)=C=\mathrm{constant}.
\end{equation}
(Primes ${}'$ stand for $d/ds$.)
If $C\neq0$ and
\begin{equation}
  \int^\infty \abs{f(s)-C}ds<\infty,
\end{equation}
then the equation \reff{eq:ode1} has a set of fundamental solutions
$\brace{e^{\pm i\sqrt C s}(1+o(1))}$ when $C>0$, or $\brace{e^{\pm
    \sqrt{\abs{C}} s}(1+o(1))}$ when $C<0$. If $C=0$ and
\begin{equation}
  \int^\infty s\abs{f(s)}ds<\infty,
\end{equation}
then the equation \reff{eq:ode1} has a set of fundamental solutions
$\brace{1+o(1), s(1+o(1))}$.
\end{theorem}

With the aid of this theorem we can prove the following.
\begin{theorem}[Asymptotic solutions for the generic case]
  \label{th:bIIIgenasym}
  Assume that the fiber eigenvalue $m$ does not vanish,
  $m\neq0$. Then, the wave equations \reff{eq:waveeqs} for the
  gauge-invariant variables \reff{eq:defQs} possess the following
  fundamental solutions
  \begin{equation}
    \begin{split}
  & \Q A(t) =\brace{t\,e^{\pm i (mt+2k\log t)}(1+o(1))}. \\
  &
  (\mathrm{A}=\mathrm{E},\mathrm{O},\mathrm{H},\text{or } \mathrm{T})
    \end{split}
  \end{equation}
\end{theorem}
\begin{proof}
  Define a new time coordinate $s$ by
  \begin{equation}
    \frac{ds}{dt}=\sigma(t)\equiv 1+\frac{2k}{t}+O(\frac{1}{t^2}).
  \end{equation}
  Rewrite the wave equation \reff{eq:waveeqs} in terms of $s$. Then, by
  taking
  \begin{equation}
    a_\mathrm{A}=\rcp{\sqrt{\sigma}}e^{-\rcp2 \int f_\mathrm{A}dt},
  \end{equation}
  and putting
    $Q_\mathrm{A}=a_\mathrm{A} X_\mathrm{A},$
  one has an alternative equation
  $X_\mathrm{A}''+W_\mathrm{A}(s)X_\mathrm{A}=0$. It is easy to check
    that $ \int^\infty \abs{W_\mathrm{A}(s)-m^2}ds<\infty $.
    Therefore, from the previous theorem, $X_A$ has the fundamental
    solutions $e^{\pm i m s}(1+o(1))$, from which the claim follows.
\end{proof}

Similarly, we can prove the following for the $m=0$ case. This case is
called the $U(1)$-symmetric case, because this kind of perturbation
keeps the $U(1)$-symmetry of the background along the $S^1$ fibers.
\begin{theorem}[Asymptotic solutions for the $U(1)$-symmetric case]
  \label{th:bIIIU1asym}
  Assume $m=0$. Define $\rho_\lambda= \lambda^2-1/4$, for
  convenience. Then, the wave equations \reff{eq:waveeqs} for the
  gauge-invariant variables \reff{eq:defQs} possess the following
  fundamental solutions: \par
    For the even perturbations:
  \begin{equation}
    \label{eq:QasymU1E}
  \Q{E}(t)=
  \begin{cases}
    \brace{\rcp{\sqrt{t}}e^{\pm i\sqrt{\rho_\lambda}\, \log t}
      \paren{ 1+o(1)}} & \quad
    (\lambda^2>\rcp4) \\
    \brace{\rcp{\sqrt{t}} (1+o(1)), \frac{\log t}{\sqrt{t}} (1+o(1))}
      & \quad
      (\lambda^2=\rcp4) \\
    \brace{t^{-\rcp2 \pm \sqrt{|\rho_\lambda|}}
      \paren{1+o(1)} }. & \quad
    (\lambda^2<\rcp4)
  \end{cases}
  \end{equation}
\par
For the odd perturbations:
  \begin{equation}
    \label{eq:QasymU1O}
  \Q{O}(t)=
  \begin{cases}
    \brace{{\sqrt{t}}e^{\pm i\sqrt{\rho_\lambda}\, \log t}
      \paren{ 1+o(1)}} & \quad
    (\lambda^2>\rcp4) \\
    \brace{{\sqrt{t}} (1+o(1)), \sqrt{t}\, {\log t} (1+o(1))}
      & \quad
      (\lambda^2=\rcp4) \\
    \brace{t^{\rcp2 \pm \sqrt{|\rho_\lambda|}}
      \paren{1+o(1)} }. & \quad
    (\lambda^2<\rcp4)
  \end{cases}
  \end{equation}
\end{theorem}
\begin{proof}
  Define a new time coordinate $s$ by
  \begin{equation}
    \frac{ds}{dt}=\sigma^\nt(t)\equiv \frac{1}{t}+O(\frac{1}{t^2}).
  \end{equation}
  Rewrite the wave equation \reff{eq:waveeqs} in terms of $s$. Then, by
  taking
  \begin{equation}
    a_A=\rcp{\sqrt{\sigma^\nt}}e^{-\rcp2 \int f_\mathrm{A}dt},
  \end{equation}
  and putting $Q_\mathrm{A}=a_A X_\mathrm{A},$ one has an alternative
  equation $X_\mathrm{A}''+W_\mathrm{A}(s)X_\mathrm{A}=0$. (A=E or O.)
  It is straightforward to check that
    \begin{equation}
      \int^\infty \abs{W_A(s)-\rho_\lambda}ds<\infty \quad
      (\lambda^2\neq \rcp4)
    \end{equation}
    and
    \begin{equation}
      \int^\infty s\abs{W_A(s)}ds<\infty. \quad
      (\lambda^2= \rcp4)
    \end{equation}
    Theorem \ref{th:1} therefore applies, and the claim follows after
    replacing the time coordinate back to $t$.
\end{proof}
\begin{remark}
  As seen from the last two proofs, the key point to find asymptotic
  solutions is the choice of a good new time variable (like the choice
  of $\sigma(t)$ or $\sigma^\nt(t)$ in the proofs.).
\end{remark}

Interestingly enough, in the generic case the asymptotic solutions are
common for all kinds of perturbations. Moreover, the asymptotic
solution does not depend on the base eigenvalue $\lambda$. In this
sense the asymptotic solutions are very universal.

On the other hand, for the $U(1)$-symmetric case, for each kind of
perturbations the behavior of the asymptotic solutions is divided into
three cases, depending upon the value of $\lambda$.

\subsection{Stability issue}
\label{ssec:III-5}

Although all the asymptotic solutions above show that our
gauge-invariant perturbation variables are growing in time, one cannot
say anything about stability of the solution from these results
themselves. This is because they do not take into account the fact
that the background solution is expanding. Because of the anisotropy
of the background solution, it is a very subtle question how one can
subtract this expansion effect. If the background was isotropic there
exists a natural way to normalize the perturbation variables, because
it has only one scale factor, but in an anisotropic case it is not
clear how to find a right way of normalizing the variables, especially
in a way such that the result is gauge invariant.

Nevertheless, one possible way may be to use the zero mode solution to
normalize the gauge invariant variables. By a \textit{zero mode
  solution} we mean a solution with the vanishing eigenvalues. Such a
solution represents a perturbation from a locally homogeneous solution
to another locally homogeneous solution, which is an effect we are not
interested in. So, it may make sense to use the zero mode solution to
do a normalization. Moreover, it is apparently gauge-invariant.
(This scheme is essentially equivalent to the one discussed in
\cite{YT}.)

All the zero mode (vacuum) solutions are already given in \cite{TMY},
from which one obtains the following growing rates for those solutions:
\begin{equation}
   \Q E^\nt = O(1), \quad
   \Q H^\nt = O(t), \quad
   \Q T^\nt = O(t^2).
\end{equation}
(Although there does not exist a zero mode for the odd perturbations,
we identify it with that of the harmonic ones, recalling the fact that
the harmonic perturbation system is formally equivalent to the limit
$\lambda\goes 0$ of the odd system.)

We define \textit{the stability measures by means of zero mode
  normalization} by
\begin{equation}
  \Mes A\equiv {\Q A}/{\Q A^\nt}, \quad (\text{A$=$E, O, H, or T})
\end{equation}
where we think of $\Q O^\nt=\Q H^\nt$ as noted above.
From the asymptotic solutions for $\Q A$ we immediately obtain the
following behaviors for our stability measures.
\begin{theorem}
  The stability measures by means of zero mode normalization for
  the vacuum orthogonal Bianchi III spacetime have the following
  asymptotic decaying or growing rates: \par
  Generic ($m\neq0$) case: \par
  \begin{equation}
      \abs{\Mes A} = 
      \begin{cases}
        O(t) & (\text{A$=$E}), \\
        O(1) & (\text{A$=$O or H}), \\
        O(t\inv) & (\text{A$=$T}).
      \end{cases}
  \end{equation}
(The rates are common for all eigenvalues (except $m=0$), but depend
on the kind of perturbations.)

  $U(1)$-symmetric ($m=0$) case: \par
  \begin{equation}
    \abs{\Mes A} = 
    \begin{cases}
      O(t^{-1/2}) & (\lambda^2>\rcp4) \\
      O(\frac{\log t}{\sqrt t}) & (\lambda^2=\rcp4) \\
      O(t^{-1/2+\sqrt{\abs{\rho_\lambda}}}) & (\lambda^2<\rcp4)
    \end{cases}
  \end{equation}
for A$=$E or O. (The rates are common for the even and odd kinds, but
there exists a critical eigenvalue $\lambda^2=1/4$ for each kind.)
\end{theorem}
\begin{remark}
  One can replace (without any modifications) the time variable $t$ to
  the proper time $\tau$ in the rate formulas of the above theorem.
\end{remark}

Note that we have three kinds of behaviors for the generic case;
The \TT\ perturbations are decaying, so we can say that the solution is
stable against this kind of perturbations. The odd and harmonic
perturbations approach a constant, so in this case the solution is
quasi-stable. Finally, the even perturbations are growing, so the
solution is \textit{unstable} against the even perturbations.

On the other hand all the $U(1)$-symmetric perturbations are decaying,
so we can say that the solution is stable against the $U(1)$-symmetric
perturbations. Although this means that the $U(1)$-symmetric
perturbations play no major role in the whole perturbations of our
system, the existence of the critical base eigenvalue $\lambda^2=1/4$
is worth pointing out. A similar existence of critical value has also
been reported in a nonlinear (but $U(1)$-symmetric) analysis
\cite{CBM}.

\section{Field equations on a closed Bianchi II solution}
\def\b{\beta}
\def\G#1{G_{\mathrm{#1}}}

\subsection{The background solution}
\label{ssec:II-1}

The Bianchi II (or Nil) group $\G{II}$ is the 3-dimensional Lie group
which consists of all real upper triangular matrices of the form
$\begin{pmatrix}
  1 & x& z \\ 0 & 1 & y \\ 0 &0 &1\\
\end{pmatrix}$ with the usual multiplication rule for matrices. To save
space let us write an element $\mathbf{x}\in\G{II}$ in (transposed)
vector form $\mathbf x=(x,y,z)$. Then the multiplication rule is given
by $(a,b,c)\circ (x,y,z)=(a+x,b+y,c+z+ay)$ for another $\mathbf
a=(a,b,c)\in\G{II}$. The group $\G{II}$ acts on our universal covering
spatial manifold $\tilde M=\R^3$ from the left, identifying the group
manifold $\G{II}$ with $\tilde M$. The Bianchi II algebra is generated
by the following three vectors $\xi_I$ ($I=1\sim3$)
\begin{equation}
  \xi_1=\vb x+y\vb z,\quad \xi_2=\vb y, \quad\xi_3=\vb z.
\end{equation}
which serve as Killing vectors on our background space or
spacetime. The invariant vectors $\c I$ and their dual 1-forms $\s I$
are given by
\begin{equation}
\begin{split}
      \chi_1 = \vb x,\quad \chi_2=\vb y+x\vb z, \quad\chi_3=\vb z, \\
      \s1 = dx,\quad \s2=dy, \quad \s3=dz-x dy.
\end{split}
\end{equation}

Like the Bianchi type III solution the Bianchi type II vacuum solution
can be expressed as
\begin{equation}
  \label{eq:bIIsol}
  (M\times\R,\dg ab)=(\tilde M\times\R,\udg ab)/\Gamma,
\end{equation}
using the universal cover solution $(\tilde M\times\R,\udg ab)$ and a
covering map $\Gamma$ acting on it.  The spatial manifold $M$ here
is, for definiteness, specified to be the ``circle bundle over the
torus ($T^2$) with Euler number $e=1$.''  (See, e.g., \cite{HemSco}.)
The fundamental group can be represented in the standard notation as
\begin{equation}
  \label{eq:pi1}
  \pi_1(M)=\angl{g_1,g_2,g_3;[g_1,g_2]=g_3,[g_1,g_3]=1,[g_2,g_3]=1},
\end{equation}
where the brackets stand for commutators, $[a,b]\equiv aba\inv b\inv$.
The vacuum metric $\udg ab$ on the universal cover is given
below. $\Gamma\subset G_\mathrm{II}$ is a discrete subgroup of the
Bianchi II group $G_\mathrm{II}$, which acts on the universal covering
spacetime $(\tilde M\times\R,\udg ab)$ from the left in the way that
keeps the natural homogeneous slicings. Some details for the
compactification are given in the next subsection. The metric $\dg ab$
on $M\times\R$ is the induced metric from $\udg ab$.

The exact vacuum solution $\udg ab$ is given \cite{Taub} by
\begin{equation}
    ds^2=-N^2(t)dt^2+q_1(t)(\s1)^2+q_2(t)(\s2)^2+q_3(t)(\s3)^2,
\end{equation}
where
\begin{equation}
  \label{eq:AQ}
  N^2= 1+\b^2t^{4p_3},\quad
  q_1=t^{2p_1}N^2,\quad
  q_2=t^{2p_2}N^2,\quad
  q_3=16p_3^2\b^2t^{2p_3}/N^2.
\end{equation}
$p_i(i=1,2,3)$ and $\b$ are constant parameters such that $\b>0$,
$p_3\neq0$, and
\begin{equation}
  \Sigma p_i=\Sigma p_i^2=1.
\end{equation}
When $p_1=p_2$, or equivalently when $q_1(t)=q_2(t)$, the solution is
said to be LRS as in the Bianchi III case. While there seem two
possible such cases $(p_1,p_2,p_3)= (0,0,1)$ and $(2/3,2/3,-1/3)$,
these two solutions represent the same one-parameter solution. When we
consider a LRS solution, we may want to take
$(p_1,p_2,p_3)=(2/3,2/3,-1/3)$, since the time coordinate $t$ in this
solution approaches the proper time at future infinity, which is more
favorable for comparisons with other models.

As shown in \cite{TKH}, $\Gamma$ is a four-parameter group. Our
spatially closed solution \reff{eq:bIIsol} therefore comprises a six
parameter solution (since the universal cover has, as we have seen,
two independent parameters).

Notice the fact that $\Gamma$ is a subgroup of the Bianchi II group
$G_\mathrm{II}$, not necessarily a subgroup of a larger group like in
the Bianchi III case. This in particular means that \textit{each}
invariant vector $\c I$ is well defined not only on the universal
cover $\tilde M$ but also on the compactified manifold $M$ (i.e., the
induced vectors $\pi^*\c I$ for the covering map $\pi:\tilde M\goes M$
are well defined on $M$. We omit  $\pi^*$ for simplicity,
though). This is because they are invariant under the action of
$G_\mathrm{II}$, and so are under $\Gamma\subset G_\mathrm{II}$. We
will see that because of this property the invariant vectors $\c I$
play a central role in developing calculus concerning mode
expansions. This significance of the invariant vectors is common for
all Bianchi class A models.

\subsection{Compactifications and eigenfunctions}
\label{ssec:II-2}

\def\N{\mathcal{N}^3}
\def\dq#1#2{q_{#1#2}}
\def\bdq#1#2{{\bar q}_{#1#2}}
\def\dqstd#1#2{q_{#1#2}^{(0)}}
\def\u{u}
\def\v{v}
\def\uv{\u\v}
\def\barA{\bar A}
\def\teiA{\u}
\def\teiB{\delta}
\def\teiC{\v}
\def\teiD{\uv}

Consider an arbitrary locally homogeneous closed 3-manifold $(M,\dq
ab)$ with $\pi_1(M)$ given by \reff{eq:pi1}. Such a Riemannian
manifold is isometric to the quotient $\N/A$, where
$\N=(\R^3,e^{2\alpha}\dqstd ab)$ is the universal cover with a metric
that is conformal to the Bianchi II standard metric
$\dqstd{}{}=(\s1)^2+(\s2)^2+(\s3)^2$. $e^{2\alpha}$ is a constant
scale factor. $A\subset\Isom\N$ is a three parameter infinite group
generated by three generators $\mathbf{a}_i\in\G{II}\subset\Isom\N$
($i=1\sim3$), which are of the form (cf. \cite{KTH})
\begin{equation}
  \mathbf{a}_1=(\u,\delta,0),\quad
  \mathbf{a}_2=(0,2\pi\v,0),\quad
  \mathbf{a}_3=(0,0,2\pi\uv).
\end{equation}
We denote $A=\brace{\mathbf{a}_1,\mathbf{a}_2,\mathbf{a}_3}$. The
three parameters $\u$, $\v$, and $\delta$ are called the \Teich\
parameters.

Also, consider another Riemannian manifold $(\bar M,\bdq ab)$, which
can be represented as $\N/\barA$, where
$\barA=\brace{\mathbf{a}_2,\mathbf{a}_3}$. $(\bar M,\bdq ab)$ is a
covering of $(M,\dq ab)$.  $\bar M$ is homeomorphic to the direct
product $T^2\times\R$, where $T^2$ is the two-torus.

Note that the (scalar) Laplacian $\Lap$ with respect to the standard
metric $\dqstd ab$ can be expressed as
$\Lap=(\c1)^2+(\c2)^2+(\c3)^2$. This operator is apparently well
defined on both $\bar M$ and $M$, since so is each $\c I$. It is
straightforward to confirm that $\Lap$ commutes with $\c 3$. We can
therefore diagonalize eigenfunctions with respect to both $\Lap$ and
$\c3$. Consider another operator $\xi_2=\del/\del y$, which we can
find commutes with both $\Lap$ and $\c3$, so one can diagonalize the
eigenfunctions with respect to $\xi_2$, too. This operator however is
\textit{not} well defined on $M$, but on $\bar M$. Because of this
fact, it is convenient to first consider eigenfunctions on $\bar M$,
and \textit{then} construct those on $M$, taking linear combinations
of the eigenfunctions on $\bar M$.

Let us define eigenvalues $\mu$ and $\nu$ for the operators $\c3$ and
$\xi_2$ by the following relations
\begin{equation}
  \c3 \bar\phi=i\mu\bar\phi,\quad \xi_2 \bar\phi=i\nu\bar\phi,
\end{equation}
where $\bar\phi$ is an eigenfunction on $\bar M$. Also, we define
$\lambda$ by $\Lap\bar\phi=-\lambda^2\bar\phi$. From the boundary
condition $\bar \phi(\bar\Gamma\x)=\bar\phi(\x)$, it is found
$ \bar\phi=X(x)e^{i\mu z}e^{i\nu y}$,
where
\begin{equation}
  \label{eq:defmunu}
  \begin{split}
  \mu &= m/(\uv), \quad m=0,\pm1,\cdots,\pm\infty, \\
  \nu &= n/\v, \quad n=0,\pm1,\cdots, \pm\infty.
  \end{split}
\end{equation}
We call $m$ (or $\mu$) the \textit{fiber eigenvalue}, $n$ (or $\nu$)
the \textit{auxiliary eigenvalue}, $\lambda$ the \textit{total
  eigenvalue}.  The function $X(x)$ must satisfy the following
equation (that is equivalent to the harmonic oscillator Schr\"odinger
equation):
\begin{equation}
  \label{eq:X}
  \X''+({\lambda^2 -{\mu^2-({\mu x+\nu})^2}})\X=0.
\end{equation}
This equation tells us in particular that like the Bianchi III case,
one must deal with the generic ($m\neq0$) case and the
$U(1)$-symmetric ($m=0$) case separately.

Let us focus on the generic ($m\neq0$) case. One can define the
eigenfunctions on $\bar M$ as
\begin{equation}
  \barphi_{l,m,n}(\x)=
  D_l(\pm\sqrt{{\frac{2\uv}{\abs{m}}}}(\frac{m}{\uv} x+\frac{n}{\v}))e^{i\frac{m}{\uv} z}e^{i\frac{n}{\v} y},
\end{equation}
where $D_l(\zeta)=e^{-\rcp4\zeta^2}H_l(\zeta)$ is the parabolic
cylinder function defined using the Hermite function
$H_l(\zeta)=(-1)^le^{\rcp2\zeta^2}\frac{d^l}{d\zeta^l}e^{-\rcp2\zeta^2}$.
The $D_l$ factor in the above equation satisfies Eq.\reff{eq:X}. 
The plus and minus signs, respectively, correspond to $m>0$ and $m<0$
cases. The index $l$ takes non-negative integers
\begin{equation}
  l=0,1,\cdots,\infty.
\end{equation}
This is related to the total eigenvalue through $\lambda^2=\abs
\mu(2l+\abs \mu+1)$.

The eigenfunctions on $M$ are, as stated, expressed as infinite sums
of these eigenfunctions on $\bar M$. 
\begin{theorem}
  Consider the generic ($m\neq0$) modes. Using the eigenfunctions
  $\bar\phi_{l,m,n}(\x)$ on the covering $(\bar M=T^2\times\R,\bdq
  ab)$, the eigenfunctions on the closed manifold $(M,\dq ab)$ are
  represented as the infinite sum
\begin{equation}
  \label{eq:philm2}
  \phi_{l,m,n_0}(\mathbf{x}) =
  \sum_{k=-\infty}^{\infty} 
  e^{ i{\teiB}(n_0k+m\frac{k(k-1)}{2})}
  \barphi_{l,m,n_0+m k}(\mathbf{x}),
\end{equation}
where $ l = 0,1,\cdots, \infty, \quad 
  |m| = 1,2,\cdots, \infty, \quad 
  n_0 = 0,1,\cdots, |m|-1 $.
\end{theorem}
\begin{proof}
  Since the summand is well defined on $\bar M$, which means it is
  invariant under the action of $\mathbf{a}_2$ and $\mathbf{a}_3$, all
  one has to confirm is the invariance of the left hand side under the
  action of $\mathbf{a}_1$. It is easy to see the
  following transformation rule
\begin{equation}
\label{eq:g1t}
  \begin{split}
    \barphi_{l,m,n}(\mathbf{a}_1\circ \x)&=
    D_l(\pm\sqrt{\frac{2\uv}{\abs{m}}}
    (\frac{m}{\uv}(x+\teiA)+\frac{n}{\v}))
    e^{i\frac{m}{\uv}(z+\teiA y)}e^{i\frac{n}{\v}(y+\teiB)}  \\
    &=  e^{i\frac{n}{\v}\teiB} 
    D_l( \pm\sqrt{\frac{2\uv}{\abs{m}}}
    (\frac{m}{\uv}x+\frac{m+n}{\v}) )
    e^{i\frac{m}{\uv}z} e^{i\frac{m+n}{\v}y}  \\
    &=  e^{i\frac\teiB\teiC n} \barphi_{l,m,n+m}(\x).
  \end{split}
\end{equation}
From this, one can confirm the invariance
\begin{equation}
  \begin{split}
  \phi_{l,m,n_0}(\mathbf{a}_1 \circ \x)
  &=
  \sum_{k=-\infty}^{\infty} 
  e^{ i\frac{\teiB}{\teiC}(n_0k+m\frac{k(k-1)}{2})}
  e^{ i\frac{\teiB}{\teiC}(n_0+mk)}\barphi_{l,m,n_0+m (k+1)}(\x)
   \\
  &=
  \sum_{k=-\infty}^{\infty} 
  e^{ i\frac{\teiB}{\teiC}(n_0(k+1)+m\frac{k(k+1)}{2})}
  \barphi_{l,m,n_0+m (k+1)}(\x)  \\
  &=
  \phi_{l,m,n_0}(\x).
  \end{split}
\end{equation}
(The last equality follows from the replacement $k\goes k-1$.)
\end{proof}

Note that as a result of the complete compactification, the index
$n_0$ for the eigenfunction on $M$ is bounded by $m$.

These eigenfunctions have the following properties.
\begin{theorem}
  For the eigenfunctions for the generic ($m\neq0$) modes, the
  following is fulfilled (indices $m$ and $n_0$ are for simplicity
  suppressed):
\begin{equation}
  \label{eq:relc}
  \begin{split}
  \chi_1\phi_{l} &= -\sqrt{\frac
  {\abs \mu}{2}}\bra{\phi_{l+1}-l\phi_{l-1}}, \\
  \chi_2\phi_{l} &= \pm i
  \sqrt{\frac{\abs \mu}{2}}\bra{\phi_{l+1}+l\phi_{l-1}}, \\
  \chi_3\phi_{l} &= i\mu\phi_{l},
  \end{split}
\end{equation}
where $\mu$ is defined in Eq.\reff{eq:defmunu} and the plus and minus
signs correspond, respectively, to $m>0$ and $m<0$.
\end{theorem}
\begin{proof}
  Use
  $D_l'(\zeta)=-\rcp2(D_{l+1}(\zeta)-lD_{l-1}(\zeta))$.
  The remaining task is a straightforward computation.
\end{proof}
\begin{remark}
  These relations can also be derived in a purely algebraic
  manner.\footnote{In fact, defining $A_1=\c1+i\c2$ and
    $A_2=\c1-i\c2$, one can check commutation relations
  \begin{equation}
    [\Lap,A_1]=2iA_1\c3,\quad  [\Lap,A_2]=-2iA_2\c3,
  \end{equation}
  which tell us, e.g., that  
  \begin{equation}
    \Lap
  A_1\phi=([\Lap,A_1]+A_1\Lap)\phi=(2iA_1\chi_3+A_1\Lap)\phi=
  -(2\mu+\lambda^2)A_1\phi.
  \end{equation}
  This means that $A_1\phi$ is an eigenfunction for eigenvalue
  $\lambda'^2=\lambda^2+2\mu$, and the operator $A_1$ acts as an
  increment (decrement) operator for $m>0$ ($m<0$). Similarly, $A_2$
  is a decrement (increment) operator with $\lambda'^2=\lambda^2-2\mu$
  for $m>0$ ($m<0$). Using these relations and following a similar
  argument for, e.g., the $SO(3)$ case to determine the ground state,
  one reaches at the following relations
  \begin{equation}
    \begin{split}
    A_1\phi_{l,m} &=-\sqrt{2 \mu}\,\phi_{l+1,m},\quad
    A_2\phi_{l,m}=\sqrt{2 \mu}\,l\,\phi_{l-1,m}, 
    \quad \text{for $m>0$} \\
    A_1\phi_{l,m} &=\sqrt{2 \abs \mu}\,l\,\phi_{l-1,m},\quad
    A_2\phi_{l,m}=-\sqrt{2 \abs \mu}\,\phi_{l+1,m},
    \quad \text{for $m<0$}
    \end{split}
  \end{equation}
  which are equivalent to the relations claimed.}
\end{remark}

\subsection{Scalar field}
\label{ssec:II-3}

As a direct application it is of great interest to see the
\textit{scalar field equation} on our Bianchi II background. For
simplicity, let us consider a background solution where $\Gam=A$,
i.e., $\Gamma$ is specified by three parameters as in the form of
$A$. We use the symbols $\u$, $\v$ and $\delta$ as in the $A$. These
parameters are constants throughout the spacetime. (This is not to say
the \Teich\ parameters for a Cauchy surface are constants of
motion. \cite{TKH}) In such a case, we can use the eigenfunctions
obtained in the previous section without any modifications. We expand
the scalar field $\Phi$ as follows:
\begin{equation}
  \Phi(t,\mathbf{x})=\sum_{l,m,n_0}
  a_{l,m,n_0}(t)\phi_{l,m,n_0}(\mathbf{x}).
\end{equation}
Since the operators $\c I$ do not change $m$ and $n_0$ when acting on
$\phi_{l,m,n_0}$, we suppress these subscripts and write simply $a_l$
and $\phi_l$, as far as no confusion occurs.

The scalar field equation $0= \ug ab\nabla_a\nabla_b\Phi$ on our
background reduces to
\begin{equation}
  \label{eq:waveII}
  0 = \frac{-1}{\sqrt{-g}}(\sqrt{-g}N^{-2}\dot\Phi)\dot{}
  + (q_1\inv(\chi_1)^2+q_2\inv(\chi_2)^2+q_3\inv(\chi_3)^2)\Phi,
\end{equation}
where a dot stands for $d/dt$. Using the background spacetime solution
for Bianchi II and the $\chi$-relations, we obtain the
following wave equations:
\begin{equation}
  \label{eq:wIIfora}
  \ddot a_l+\rcp t\dot a_l+\mu f(t) a_l=
  \frac \mu 2 (t^{-2p_1}-t^{-2p_2})[a_{l-2}+(l+2)(l+1)a_{l+2}],
\end{equation}
where
\begin{equation}
  f(t)\equiv \frac{2l+1}2(t^{-2p_1}+t^{-2p_2})
    +\frac{\mu}{16(p_3)^2\b^2}(1+\b^2t^{4p_3})^2t^{-2p_3}.
\end{equation}
(In Eq.\reff{eq:wIIfora}, $a_{l<0}$ should be considered as zero.)

A notable property of these equations is that they comprise a set of
infinitely coupled equations, unless the solution is LRS. These
couplings between different modes are an unavoidable feature for the
field equations on a non-LRS spacetime solution, because for those
backgrounds the mode functions defined with respect to a standard
metric are no longer eigenfunctions with respect to the Laplacian for
each spatial section. That is, those eigenfunctions defined on the
standard Riemannian manifold do not remain eigenfunctions in the
course of the anisotropic expansion. One could define another
``time-dependent'' eigenfunctions to get around this difficulty, but
in that case time-derivatives of the eigenfunctions appear in the
field equation, so in any case the complexities caused by anisotropic
expansion are unavoidable.

The above wave equation gives a closed equation when the background is
LRS. In this case we can again find asymptotic solutions:
\begin{theorem}
  On the LRS Bianchi II vacuum solution with $p_1=p_2=2/3$ and
  $p_3=-1/3$, the scalar field equation \reff{eq:wIIfora} for the
  generic mode has the following fundamental solutions as
  $t\goes\infty$:
\begin{equation}
  a_l(t)= \brace{t^{-\frac23}
    e^{\pm i\frac{3\mu}{4\b}(\frac34t^{4/3}+\b^2\log t)}(1+o(1))}.
\end{equation}
\end{theorem}
\begin{proof}
  As remarked in the previous section, it is a suitable choice
  of new time variable (denoted $s$) that is essential to have an
  asymptotic solution. An appropriate choice for the present equation
  is
  \begin{equation}
    \frac{ds}{dt}= \frac{3}{4}\paren{\frac{t^{1/3}}{\b}+\frac{\b}{t}}.
  \end{equation}
  One can show the claim following the same procedure as in
  \textsc{Theorem} \ref{th:bIIIgenasym} or \ref{th:bIIIU1asym}.
\end{proof}

\subsection{Vectors and electromagnetic field}
\label{ssec:II-4}

Let us consider how we can perform separation of variable for
\textit{vector fields}. An advantage that our Bianchi II model has and
is common for other class A spatially closed Bianchi models is that in
principle, we do \textit{not} have to define suitable eigenvectors for
this purpose. This is because there exists a set of well-defined
independent three invariant vectors $\c I$ on the compactified
manifold $M$. Any vectors on $M$ can be expanded with respect to these
invariant vectors (i.e., one can consider the components with respect
to the frame $\brace{\c I}$) in a group invariant way and so all one
has to do is to mode-expand the components using the already defined
eigenfunctions $\phi_{l,m,n_0}$. However, this straightforward
expansion is found not to be very useful for a successful analysis.

A nicer way is to use an analogy with Regge-Wheeler's spherically
symmetric case or one with Bianchi III hyperbolically symmetric case,
where one can take advantage of the spherically or hyperbolically
symmetric planes. In those cases, one is able to define so-called the
`even' vectors and `odd' vectors on the symmetric plane (as seen in
the previous section for the Bianchi III case), and the even and odd
part equations for a linear vector field equation decouple from each
other.

Although there do not exist similar symmetric planes for the Bianchi
II case, we can, instead, consider the ``plane field'' spanned by
$\chi_1$ and $\chi_2$. This plane field is not integrable, because the
commutator $[\c1,\c2]=\c3$ is independent from $\c1$ and
$\c2$. Nevertheless, the plane field can play a special role, because
the metric can be LRS only when $q_1=q_2$.

We define the area two-form $\varepsilon$ of the plane field by $
\varepsilon_{ab}=2\ss1{[a}\ss2{b]}$.  We may define the even vectors
by $  S_a=(\c1\phi_l)\ss1a+(\c2\phi_l)\ss2a$, and odd vectors by $
V_a=i\varepsilon_a{}^b\del_b\phi_l
=i((\c2\phi_l)\ss1a-(\c1\phi_l)\ss2a)$.  (To raise an index for
$\varepsilon_{ab}$ we use the standard Bianchi II metric.)  We also
need orthogonal vectors $S_a^{\perp}=(\c3\phi_l)\ss3a$ and time-like
vectors $S_a^\mathrm{T}=N\inv\phi_l(dt)_a$, which may be considered to
belong to the even part. These four kinds of vectors complete our set
of eigenvectors.

Because of the fact that the commutator of $\c1$ and $\c2$ does not
belong to the tangent space spanned by $\c1$ and $\c2$, the even and
odd parts defined this way do \textit{not} decouple from each other,
even when the background is LRS. Nonetheless, the use of even and odd
vectors is found useful, especially when the background is LRS, since
in that case each mode is found to decouple from the others. (Do not
confuse couplings between the even and odd parts and ones between
modes.) When the spacetime is not LRS, however, we will have couplings
between different modes due to the anisotropic expansion.

Using these mode vectors, we can expand the \textit{vector potential}
for an electromagnetic field as follows:
\begin{equation}
  A_a=\sum_{l,m,n_0} \alpha_0(t)
  S_a^\mathrm{T}+\alpha_1(t)S_a+\alpha_2(t)V_a+\alpha_3(t) S_a^{\perp}.
\end{equation}
The four kinds of functions $\alpha_i(t)$ ($i=0\sim3$) serve as the
field variables. Let us consider the \textit{LRS case} below, on which
background, as mentioned above, couplings between different modes
vanish.

The quantities we are most interested in are the
($U(1)$-)gauge-invariant variables, which can be easily found by
looking at components of the field strength
$F_{ab}=\del_aA_b-\del_bA_a$. We obtain the following four independent
gauge invariant variables:
\begin{equation}
  Q_1=\alpha_1-\alpha_3,\;
  Q_2=\alpha_2,\;
  P_1=\dot\alpha_1-\alpha_0,\;
  P_2=\dot\alpha_2.
\end{equation}
Although function $P_3\equiv \dot\alpha_3-\alpha_0$ is also gauge
invariant, it is found that it can be (consistently) solved with the
others, due to the constraint part of Maxwell's equation. The
evolution equations for the LRS background are found as follows:
\begin{equation}
  \begin{split}
    \dot Q_1 &= P_1+\rcp \mu\frac{q_3}{q_1}((2l+1)P_1+P_2),\quad
    \dot Q_2 = P_2, \\
    \dot P_1 &= \paren{\frac{\dot N}{N}-\rcp2\frac{\dot q_3}{q_3}}P_1
    -\mu^2\frac{N^2}{q_3}Q_1, \\
    \dot P_2 &= \paren{\frac{\dot N}{N}
      -\rcp2\frac{\dot q_3}{q_3}}P_2
    -\mu^2\frac{N^2}{q_3}Q_2
    -\mu\frac{N^2}{q_1}(Q_1+(2l+1)Q_2).
  \end{split}
\end{equation}

To tackle the gravitational perturbation equation we need to
generalize our framework so that it is applicable to systems with
symmetric tensors. Again, thanks to the well-defined invariant vectors
$\c I$, separations of variable for the tensors are straightforward in
principle. However, because of the complexity of the linearized
Einstein equation, to make our analysis successful we will need to
find a well-arranged set of base mode tensors like the even and odd
vectors used above.

\bibliographystyle{amsalpha}

\begin{thebibliography}{AAA}

\bibitem [CBM]{CBM} Y. Choquet-Bruhat, and V. Moncrief, \textit{Future
    global in time Einsteinian spacetimes with U(1) isometry group}
  Ann. Henri Poincar\'e \textbf{2} (2001) 1007--1064;  \textit{Non
    linear stability of einsteinian spacetimes with U(1) isometry
    group}, preprint (gr-qc/0302021).

\bibitem [CH]{CH} L. Cesari, \textit{Asymptotic behavior and stability
    problems in ordinary differential equations}, Springer, 3rd ed.
  (1971); P. Hartman, \textit{Ordinary differential equations}, John
  Wiley (1964).
  
\bibitem [EM]{EM} G.F.R. Ellis, and M.A.H. MacCallum, \textit{A
class of homogeneous cosmological models},
Commun. Math. Phys. \textbf{12} (1969) 108--141.
  
\bibitem[FM]{FM} A.E. Fischer, and V. Moncrief, \textit{The reduced
    Einstein equations and the conformal volume collapse of
    3-manifolds}, Class. Quantum Grav. \textbf{18} (2001) 4493--4515;
  and reference therein.
  
\bibitem[HS]{HemSco} J.~Hempel, \textit{3-manifolds}, Ann. of
  Math. Studies 86, Princeton University Press (1976); P.~Scott,
  \textit{The geometry of 3-manifolds}, Bull. London Math. Soc. {\bf
    15} (1983) 401--487.
  
\bibitem[KTH]{KTH} T.~Koike, M.~Tanimoto, and A.~Hosoya,
  \textit{Compact Homogeneous Universes}, J. Math. Phys. {\bf 35}
  (1994) 4855--4888.
  
\bibitem[TAU]{Taub} A. H. Taub, \textit{Empty space-times admitting a
    three-parameter group of motions}, Ann. Math. \textbf{53} (1952)
  472--490.
  
\bibitem[TKH]{TKH} M.~Tanimoto, T.~Koike, and A.~Hosoya,
  \textit{Dynamics of compact homogeneous universes},
  J. Math. Phys. {\bf 38} (1997) 350--368; \textit{Hamiltonian
    structures for compact homogeneous universes}, J. Math. Phys. {\bf
    38} (1997) 6560--6577.

\bibitem[TMY]{TMY} M.~Tanimoto, V.~Moncrief, and K.~Yasuno,
  \textit{Perturbations of spatially closed Bianchi III spacetimes},
  Class. Quantum Grav. \textbf{20} (2003) 1879--1927.

\bibitem[WE]{WE}
J. Wainwright, and G.F.R. Ellis,
\textit{Dynamical Systems in Cosmology\/},
Cambridge University Press (1997).

\bibitem[YT]{YT} K.~Yasuno, and M.~Tanimoto, \textit{Asymptotic
    behavior of linearly perturbed spatially closed Bianchi III vacuum
    solutions}, in preparation.
  
\bibitem[YO]{Yo} J.W.~York~Jr., \textit{Conformally invariant
    orthogonal decomposition of symmetric tensors on Riemannian
    manifolds and the initial-value problem of general relativity},
  J. Math. Phys. {\bf 14} (1973) 456--464.

\end{thebibliography}

\vspace{4em}

\sf

{\large Errata to the published version}

This article will be published in \textit{Contemporary Mathematics} of
American Mathematical Society. The following is corrections to that
published version. (Unfortunately, authors to this publication are not
given a chance to make corrections after the final submission.) All
the corrections have already been incorporated in the present
electronic version, so there is no need to make any change for this
article.

\textbf{1}. The line below Eq.(3.14) should read 

``where $ l = 0,1,\cdots, \infty,
\quad |m| = 1,2,\cdots, \infty, \quad n_0 = 0,1,\cdots, |m|-1 $.''

\textbf{2}. Eq.(2.2) and Eq.(3.1) should be swapped.

\end{document}